\documentclass[12pt]{article}

\usepackage{graphicx}

\usepackage{latexsym,cmmib57,amssymb}
\usepackage{amsmath,amsthm}
\usepackage[psamsfonts]{eucal}
\usepackage{url}

\newcommand{\citep}[1]{\cite{#1}}
\newtheorem{theorem}{Theorem}

\newtheorem{corollary}[theorem]{Corollary}
\newtheorem{lemma}[theorem]{Lemma}

\newcommand{\condgen}[6]{{#1}#2 #5 #3 #6 #4}
\newcommand{\bbrd}[1]{\mbox{\rm{I}\kern-.1667em{#1}}}

\newcommand{\PROB}{\mathbb{P}}

\newcommand{\Probcmd}[2]{\condgen{\PROB}{\Bigl\{}{\Bigm|}{\Bigr\}}{#1}{#2}}

\newsavebox{\fmbox}
\newenvironment{fmpage}[1]
        {\begin{lrbox}{\fmbox}\begin{minipage}{#1}}
        {\end{minipage}\end{lrbox}\fbox{\usebox{\fmbox}}}

\newcounter{algocnt}
\newenvironment{algolist}[1]{%
    \begin{list}{\thealgocnt}
    {\parsep 0in\usecounter{algocnt}\setcounter{algocnt}{0}\renewcommand{\thealgocnt}{{#1}\arabic{algocnt}}
    \setlength{\rightmargin}{0in}
    \settowidth{\leftmargin}{{#1}999}\addtolength{\leftmargin}{\labelsep}}}{\end{list}}
\newcommand{\algotab}[0]{\hspace*{1.2\labelsep}}
\newcommand{\mt}[0]{\algotab\ }
\newcommand{\mtt}[0]{\algotab\ \algotab\ }
\newcommand{\mttt}[0]{\algotab\ \algotab\ \algotab\ }
\newcommand{\mtttt}[0]{\algotab\ \algotab\ \algotab\ \algotab\ }
\newcommand{\mttttt}[0]{\algotab\ \algotab\ \algotab\ \algotab\ \algotab\ }
\newcommand{\Algo}[1]{\textsc{#1}}
\newcommand{\setto}{\leftarrow}

\newcommand{\tree}{T}
\newcommand{\nodelabel}{\xi}
\newcommand{\snodelabel}{\Xi}
\newcommand{\lrate}{\mu}
\newcommand{\drate}{\lambda}
\newcommand{\grate}{\kappa}
\newcommand{\rootdist}{\gamma}

\newcommand{\extinct}[1]{D_{#1}}
\newcommand{\prodextinct}[1]{D[#1]}

\newcommand{\ehori}[2]{H_{#1}(#2)}
\newcommand{\hori}[1]{H(#1)}
\newcommand{\ecopies}[2]{G_{#1}(#2)}
\newcommand{\copies}[1]{G(#1)}

\newcommand{\numchildren}{c}

\newcommand{\likelihoodsym}{L}
\newcommand{\likelihood}[2]{\likelihoodsym_{#1}[#2]}
\newcommand{\sectionlik}[2]{A_{#1;#2}}
\newcommand{\lineagelik}[3]{B_{#1; #2, #3}}

\newcommand{\xenosize}{\eta}
\newcommand{\parasize}{\zeta}

\newcommand{\leafset}{\mathcal{L}}
\newcommand{\leaves}[1]{\leafset(#1)}

\newcommand{\psumsym}{w}
\newcommand{\psume}[3]{\psumsym_{#1}[{#3}\vert{#2}]}
\newcommand{\psumo}[2]{\psumsym[{#2}\vert{#1}]}
\newcommand{\psumi}[2]{\psumsym^{*}[{#2}\vert {#1}]}
\newcommand{\psumei}[3]{\psumsym^{*}_{#1}[{#3}\vert{#2}]}

\newcommand{\sumsize}[1]{M[#1]}

\newcommand{\profile}{\Phi}

\begin{document}
\title{Mathematical Framework for Phylogenetic Birth-And-Death Models}
\author{Mikl\'os Cs\H{u}r\"os\thanks{%
			University of Montr\'eal, Department of Computer Science and Operations Research, Canada. E-mail: csuros AT iro.umontreal.ca} 
			\and 
		Istv\'an Mikl\'os\thanks{%
			Alfr\'ed R\'enyi Institute of Mathematics, Hungarian Academy of Sciences.}
		}
\maketitle

\begin{abstract}
A phylogenetic birth-and-death model is a probabilistic 
graphical model for a so-called phylogenetic profile, i.e., 
the size distribution for a homolog gene family at 
the terminal nodes of a phylogeny. Profile datasets are 
used in bioinformatics analyses for the inference of 
evolutionary trees, and of functional associations between gene families, 
as well as for the quantification of various processes 
guiding genome evolution. Here we describe the mathematical formalism 
for phylogenetic birth-and-death models. We also present an algorithm  
for computing the likelihood in a gain-loss-duplication model. 
\end{abstract}

For background information on 
phylogenetic birth-and-death models, see~\cite{Istvan.genecontent} 
(preprint available under \url{http://arxiv.org/abs/q-bio/0509037v1}).
Here we give a self-containg comprehensive review, concentrating on the mathematical 
results. We describe 
our new algorithm for a very general class of gain-loss-duplication 
models.

\section{Introduction}
A {\em phylogenetic birth-and-death model} formalizes the 
evolution of an organism-specific census variable along 
a phylogeny. 
The phylogeny is a rooted tree, i.e., 
a connected acyclic graph in which the edges are directed away from a 
special node designated as the tree root; the terminal nodes, or {\em leaves}, are 
bijectively labeled by the organisms. 
The model specifies edge lengths,  
as well as birth-and-death processes \citep{Ross,Kendall} acting on the edges. 
Let~$\mathcal{E}(\tree)$ denote the set of edges, and 
let~$\mathcal{V}(\tree)$ denote the node set of the tree.
Populations of identical individuals evolve along the tree 
from the root towards the leaves by Galton-Watson processes. 
At non-leaf nodes of the tree, populations are 
instantaneously copied to evolve independently along the 
adjoining descendant edges. 
Let the random variable~$\nodelabel(x)\in\mathbb{N}=\{0,1,2,\dotsc\}$ denote the population count 
at every node~$x\in \mathcal{V}(\tree)$.
Every edge~$xy\in \mathcal{E}(\tree)$ is characterized by a loss 
rate~$\lrate_{xy}$, a duplication rate~$\drate_{xy}$ and a gain rate~$\grate_{xy}$.
If~$\bigl(X(t)\colon t\ge 0\bigr)$ is a linear birth-and-death process \citep{Kendall,Takacs} with these rate parameters, then 
\[
\Probcmd{\nodelabel(y)=m}{\nodelabel(x)=n} = \Probcmd{X(t_{xy})=m}{X(0)=n}, 
\]
where~$t_{xy}>0$ is the edge length, which defines the time interval during which the birth-and-death process runs. 
The joint distribution of~$(\nodelabel(x)\colon x\in \mathcal{V}(\tree))$ is 
determined by the phylogeny, the edge lengths and rates, along with the distribution at the root~$\rho$, 
denoted as $\rootdist(n)=\PROB\{\nodelabel(\rho) = n\}$. 
Specifically, for all set of node census values $\bigl(n_x\colon x\in \mathcal{V}(\tree)\bigr)$, 
\begin{equation}\label{eq:likelihood.allcounts}
\PROB\Bigl\{\forall x\in \mathcal{V}(\tree)\colon \nodelabel(x) = n_x\Bigr\}
	= \rootdist(n_{\rho}) \prod_{xy\in \mathcal{E}(\tree)} \psume{xy}{n_x}{n_y} 
\end{equation}
where $ \psume{xy}{n}{m} =\Probcmd{X(t_{xy})=m}{X(0)=n}$ denotes the transition probability
on the edge~$xy$ for the Markov process operating there. 

It is assumed that one can observe the population counts at
the terminal nodes (i.e., leaves), but not at the inner
nodes of the phylogeny. Since individuals are considered
identical, we are also ignorant of the ancestral
relationships between individuals within and across
populations. The population counts at the leaves form a {\em
phylogenetic profile}. Our central problem is to compute the
likelihood of a profile, i.e., the probability of the
observed counts for fixed model parameters.

The transient distribution of linear birth-and-death processes is well-characterized
\citep{KarlinMcgregor,Kendall,Takacs}, as shown in Table~\ref{tbl:bd}. 
Table~\ref{tbl:bd} precisely states the distribution of xenolog and inparalog group sizes. 

\begin{table}\footnotesize
\begin{center}
\begin{tabular}{llll}
\hline
\textbf{Case} & \textbf{Condition} & \textbf{Transient distribution} & \textbf{Group}\\
\hline
\textbf{GLD} & $\grate>0, \drate>0$ & $\Probcmd{X(t)=n}{X(0)=0} = \mathsf{NegativeBinomial}(n; \theta, q)$ & xenolog \\
\textbf{GL}  & $\grate>0, \drate=0$  & $\Probcmd{X(t)=n}{X(0)=0} = \mathsf{Poisson}(n; r)$ & xenolog\\
\textbf{DL}  & $\grate=0,  \drate>0$ & $\Probcmd{X(t)=n}{X(0)=1} = \mathsf{ShiftedGeometric}(n; p,q)$ & inparalog \\\
\textbf{PL}  & $\grate=0, \drate=0$ & $\Probcmd{X(t)=n}{X(0)=1} = \mathsf{Bernoulli}(n; 1-p)$ & inparalog\\
\hline
\end{tabular}
\end{center}

\paragraph{Parameters:}
\begin{align*}
\theta & = \frac{\grate}{\drate} \qquad r =  \grate \frac{1-e^{-\lrate t}}{\lrate}\\*
p & = 
	\frac{\lrate-\lrate e^{-(\lrate-\drate)t}}{\lrate-\drate e^{-(\lrate-\drate)t}} \quad\text{and}\quad
	q = \frac{\drate-\drate e^{-(\lrate-\drate)t}}{\lrate-\drate e^{-(\lrate-\drate)t}} 
	& \text{if $\drate \ne \lrate$,}\\
p & = q = \frac{\drate t}{1+\drate{}t} & \text{if $\drate = \lrate$.} 
\end{align*}

\paragraph{Distributions:}
\begin{align*}
\mathsf{NegativeBinomial}(n; \theta, q) 
	& = \begin{cases}
			 (1-q)^{\theta} & \text{if $n=0$}\\*
			\frac{\theta (\theta+1)\dotsm (\theta+n-1)}{n!}(1-q)^{\theta} q^{n} & \text{if $n>0$}
		\end{cases}\\
\mathsf{ShiftedGeometric}(n; p, q) 
	& = \begin{cases}
		p & \text{if $n=0$;}\\*
		(1-p)(1-q) & \text{if $n=1$}\\*
		(1-p)(1-q) q^{n-1} & \text{if $n>1$.}
		\end{cases}\\
\mathsf{Poisson}(n; r) 
	& =  e^{-r} \frac{r^n}{n!}\\
\mathsf{Bernoulli}(n; 1-p) 
	& = \mathsf{ShiftedGeometric}(n; p, 0) = \begin{cases} 
			p & \text{if $n=0$}\\*
			1-p & \text{if $n=1$}. 
		\end{cases}
\end{align*}
\caption[Transient behavior of linear birth-and-death processes]{Transient behavior of linear birth-and-death processes with loss rate~$\lrate>0$, gain rate $\grate$ and duplication rate~$\drate$:
	gain-loss-duplication (\textbf{GLD}),
	gain-loss (\textbf{GL}), duplication-loss (\textbf{DL}) and pure-loss (\textbf{PL}) models.
	The last column of the table shows the relevant group for computing transition probabilities 
	in a phylogenetic birth-and-death model.
	For the meaning of xenolog and inparalog groups, see the main text. 
	}\label{tbl:bd}
\end{table}

The distribution of population counts can be 
obtained analytically from the constituent distributions of Table~\ref{tbl:bd}, as shown by the following lemma. 
\begin{lemma}\label{lm:psum}
Let~$\bigl(\parasize_i\colon i=1,2,\dotsc\bigr)$ be independent random variables 
that have identical, shifted geometric distributions 
with parameters~$p$ and~$q$. Let~$\xenosize$ be a discrete nonnegative random variable 
that is independent from $\parasize_i$, with probability mass function 
$\PROB\{\xenosize=m\} = \hori{m}$.
Define~$\psumo{n}{m} = \PROB\{\xenosize+\sum_{i=1}^n \parasize_i=m\}$
for all $m,n\ge 0$, and $\psumi{n}{m} = \PROB\{\xenosize+\sum_{i=1}^n \parasize_i=m; \forall \parasize_i>0\}$
for all $m\ge n\ge 0$. 
These values can be expressed recursively as follows.
\begin{subequations}\label{eq:sumo}
\begin{align}
\psumo{0}{m} & = \hori{m} & \{m\ge 0\}\label{eq:sumo.0m}\\
\psumo{n}{0} & = p\cdot \psumo{n-1}{0} & \{n>0\}\label{eq:sumo.n0}\\
\psumo{n}{1} & = p\cdot \psumo{n-1}{1} + (1-p)(1-q) \cdot\psumo{n-1}{0}
	& \{n>0\}\label{eq:sumo.n1}\\
\psumo{n}{m} & =
	\begin{aligned}[t]
	  & q \cdot\psumo{n}{m-1} \\
	&+ (1-p-q) \cdot\psumo{n-1}{m-1} \\
	&+ p \cdot\psumo{n-1}{m} 
	\end{aligned} & \{n>0,m>1\} \label{eq:sumo.nm}
\end{align}
\end{subequations}
Furthermore,
\begin{subequations}\label{eq:sumi}
\begin{align}
\psumi{0}{m} & = \hori{m} & \{m\ge 0\} \label{eq:sumi.0m}\\
\psumi{n}{n} & = (1-p)(1-q) \cdot\psumi{n-1}{n-1} & \{n>0\}\label{eq:sumi.nn}\\*
\psumi{n}{m} & = 
	\begin{aligned}[t]
		& q \cdot\psumi{n}{m-1}\\*
		&+(1-p)(1-q)\cdot \psumi{n-1}{m-1}
	\end{aligned} & \{m>n>0\}\label{eq:sumi.nm}
\end{align}
\end{subequations}
\end{lemma}

For every edge~$xy$, Equation~\eqref{eq:sumo} provides the 
transition probabilities~$\psume{xy}{n}{m}=\psumo{n}{m}$
in~\eqref{eq:likelihood.allcounts}, when~$p$, $q$ and $\hori{m}$ 
are taken from Table~\ref{tbl:bd} for the process operating on the edge~$xy$.
Equation~\eqref{eq:sumi} is used below in our formulas.

\section{Surviving lineages}
A key factor in inferring the likelihood formulas is the probability that 
a given individual at a tree node~$x$ has no descendants at 
the leaves within the subtree rooted at~$x$. 
The corresponding {\em extinction probability} is denoted by~$\extinct{x}$. 
An individual at node~$x$ is referred to as {\em surviving} if it has at least one progeny 
at the leaves descending from~$x$. 
Let~$\snodelabel(x)$ denote the number of surviving individuals at each node~$x$. 
The distribution of~$\snodelabel(x)$ can be related to that of~$\nodelabel(x)$ by
\begin{equation}\label{eq:snodelabel.calc}
\PROB\bigl\{\snodelabel(x) = m\bigr\} 
	= \sum_{i=0}^\infty \binom{m+i}{i} \extinct{x} ^i (1-\extinct{x})^m \PROB\bigl\{\nodelabel(x) = m+i\bigr\}. 
\end{equation}

The next two lemmas 
characterize the number of surviving xenologs and inparalogs: they follow the same 
class of distributions as the total number of xenologs and inparalogs.

\begin{lemma}\label{lm:paralog.survive}
For every edge~$xy\in \mathcal{E}(\tree)$, 
let $\ecopies{y}{n}$ denote the probability that 
there are~$n$ surviving members within an inparalog group at~$y$.
Then $\ecopies{y}{n}=\mathsf{ShiftedGeometric}(n; p',q')$
with 
\[
p' = \frac{p(1-\extinct{y})+(1-q)\extinct{y}}{1-q\extinct{y}}\quad\text{and}\quad
q' = \frac{q(1-\extinct{y})}{1-q\extinct{y}}.
\]
\end{lemma}

\begin{lemma}\label{lm:xenolog.survive}
For every edge~$xy\in \mathcal{E}(\tree)$, 
let $\ehori{y}{n}$ denote the probability that
there are~$n$ xenologs at~$y$ that survive.
If $\drate_{xy}=0$, then $\ehori{y}{n}=\mathsf{Poisson}(n;r')$
where $r'=r(1-\extinct{y})$.
If $\drate_{xy}>0$, then $\ehori{y}{n}=\mathsf{NegativeBinomial}(n; \theta, q')$.
\end{lemma}

In the formulas to follow, we use 
the probabilities $\psumei{y}{n}{m}$, which apply 
Lemma~\ref{lm:psum} to surviving populations on edge~$xy$: 
$\psumei{y}{n}{m}=\psumi{n}{m}$, where the latter is defined by 
Equation~\eqref{eq:sumi} with settings $p\leftarrow p'$, $q\leftarrow q'$, 
$\hori{m}\leftarrow\ehori{x_i}{m}$ from Lemmas~\ref{lm:paralog.survive} and~\ref{lm:xenolog.survive}.

Lemma~\ref{lm:paralog.survive} provides the means to compute 
extinction probabilities in a postorder traversal of the phylogeny.
\begin{lemma}\label{lm:extinct}
If~$x$ is a leaf, then~$\extinct{x}=0$. 
Otherwise, let~$x$ be the parent of~$x_1, x_2,\dotsc, x_{\numchildren}$.
Then $\extinct{x}$ can be written as 
\begin{equation}\label{eq:extinct}
\extinct{x} = \prod_{j=1}^{\numchildren} \ecopies{x_j}{0}.
\end{equation}
\end{lemma}

\section{Conditional likelihoods}
Let~$\leaves{\tree}\subset\mathcal{V}(\tree)$ denote the set of leaf nodes.
A phylogenetic profile~$\profile$ is a function $\leaves{\tree}\mapsto\{0,1,2,\dotsc\}$, 
which are the population counts observed at the leaves. 
Define the notation $\profile(\leafset')=\bigl(\profile(x)\colon x\in\leafset'\bigr)$ for the 
partial profile within a subset~$\leafset'\subseteq\leaves{\tree}$.  
Similarly, let~$\nodelabel(\leafset')=\bigl(\nodelabel(x)\colon x\in\leafset'\bigr)$ denote the vector-valued 
random variable composed of individual population counts. 
The {\em likelihood} of~$\profile$ is the probability 
\begin{equation}\label{Meq:likelihood.def}
\likelihoodsym = \PROB\Bigl\{\nodelabel\bigl(\leaves{\tree}\bigr)=\profile\Bigr\}.  
\end{equation}
Let~$\tree_x$ denote the subtree of~$\tree$ rooted at node~$x$.
Define the {\em survival count range} $M_x$ for every node~$x\in\mathcal{V}(\tree)$ as
$M_x = \sum_{y\in\leaves{\tree_x}} \profile(y)$.  
The survival count ranges are calculated in a postorder traversal, since
\begin{equation}\label{eq:range}
M_x = 
	\begin{cases}
	\profile(x) & \text{if $x$ is a leaf}\\
	\sum_{y\in\mathsf{children}(x)} M_y & \text{otherwise.}
	\end{cases}
\end{equation}

We compute the likelihood using {\em conditional survival likelihoods} defined as 
the probability of observing the partial profile within~$\tree_x$ given
the number of surviving individuals $\snodelabel(x)$:
\[
\likelihood{x}{n} = \Probcmd{\nodelabel\bigl(\leaves{\tree_x}\bigr)=\profile\bigl(\leaves{\tree_x}\bigr)}{\snodelabel(x)=n}.
\]
For~$m>M_x$, $\likelihood{x}{m}=0$. 
For values~$m=0,1,\dotsc,M_x$, 
the conditional survival likelihoods can be computed recursively as shown in Theorem~\ref{tm:likelihood.recursion} below.
\begin{theorem}\label{tm:likelihood.recursion}
If node~$x$ is a leaf, then 
\[
\likelihood{x}{n} 
	= \begin{cases}
		0 & \text{if $n \ne \profile(x)$;}\\*
		1 & \text{if $n = \profile(x)$.}
	\end{cases}
\]
If~$x$ is an inner node with children~$x_1,\dotsc,x_{\numchildren}$, then 
	$\likelihood{x}{n}$ can be expressed using $\likelihood{x_i}{\cdot}$ 
	and auxiliary values $\sectionlik{i}{\cdot}$ and~$\lineagelik{i}{\cdot}{\cdot}$
	for $i=1,\dotsc,\numchildren$
	in the following manner.
Let $\psumei{xx_i}{s}{m}$ denote the 
transition probability in Lemma~\ref{lm:psum}, applied
to surviving individuals at~$x_i$, using the distributions~$\ehori{x_i}{\cdot}$ 
from Lemma~\ref{lm:xenolog.survive} and $\ecopies{x_i}{\cdot}$ from Lemma~\ref{lm:paralog.survive}. 
Let~$\sumsize{j}=\sum_{i=1}^j M_{x_i}$ for all $j=1,\dotsc,\numchildren$ and $\sumsize{0}=0$. 
Define also $\prodextinct{j} = \prod_{i=1}^j \ecopies{x_i}{0}$ and $\prodextinct{0}=1$.
Auxiliary values $\lineagelik{i}{t}{s}$ are defined for all $i=1,\dotsc,\numchildren$, 
$t=0,\dotsc,\sumsize{i-1}$ and $s=0,\dotsc,M_{x_i}$ as follows.
\begin{subequations}\label{eq:lineagelik}
\begin{align}
\lineagelik{i}{0}{s} & = \sum_{m=0}^{M_{x_i}} \psumei{xx_i}{s}{m} \likelihood{x_i}{m} & \{0\le s\le M_{x_i}\}\label{eq:llik.i0s}\\* 
\lineagelik{i}{t}{M_{x_i}} & = \ecopies{x_i}{0} \lineagelik{i}{t-1}{M_{x_i}} & \{0<t\le \sumsize{i-1}\}\label{eq:llik.itm}\\*
\lineagelik{i}{t}{s} & =\lineagelik{i}{t-1}{s+1} + \ecopies{x_i}{0} \lineagelik{i}{t-1}{s} 
	&	\Bigl\{\begin{smallmatrix}
			0\le s<M_{x_i}\\
			0<t\le\sumsize{i-1}
			\end{smallmatrix}\Bigr\} \label{eq:llik.its}
\end{align}
\end{subequations}
For all~$i=1,\dotsc,\numchildren$ and $n=0,\dotsc,\sumsize{i}$, define $\sectionlik{i}{n}$ as
\begin{subequations}\label{eq:recursion.section}
\begin{align}
\sectionlik{1}{n} & = (1-\prodextinct{1})^{-n} \lineagelik{1}{0}{n}; \label{eq:recursion.section.1}\\
\sectionlik{i}{n} & = \bigl(1-\prodextinct{i}\bigr)^{-n} 
	\sum_{\substack{0\le t\le M[i-1]\\0\le s\le M_{x_i}\\t+s=n}}  
	\mathsf{Binomial}(s; n, \prodextinct{i-1})
	\sectionlik{i-1}{t} \lineagelik{i}{t}{s},\label{eq:recursion.section.i}
\end{align}
\end{subequations}
where $i>1$ in~\eqref{eq:recursion.section.i}.

For all $n=0,\dotsc,M_x$, $\likelihood{x}{n} = \sectionlik{\numchildren}{n}$.
\end{theorem}

The complete likelihood is computed as 
\begin{multline}\label{eq:likelihood.infinite}
\likelihoodsym = \sum_{m=0}^{M_{\rho}} \likelihood{\rho}{m}\PROB\{\snodelabel(\rho)=m\}\\
	= \sum_{m=0}^{M_{\rho}} \likelihood{\rho}{m}\Biggl(\sum_{i=0}^\infty \rootdist(m+i) \binom{m+i}{i} \extinct{\rho}^i (1-\extinct{\rho})^m\Biggr).
\end{multline}

For some parametric distributions~$\rootdist$, the infinite sum in~\eqref{eq:likelihood.infinite}
can be replaced by a closed formula for~$\PROB\{\snodelabel(\rho)=m\}$. Theorem~\ref{tm:likelihood.finite} below considers
the stationary distributions for gain-loss-duplication and gain-loss models.
\begin{theorem}\label{tm:likelihood.finite}
For negative binomial or Poisson population distribution at the root, 
the likelihood can be expressed as shown below.
\begin{subequations}
\begin{enumerate}
\item If $\rootdist(n)=\mathsf{Poisson}(n;r)$, then
\begin{equation}\label{eq:likelihood.finite.Poisson}
\PROB\{\snodelabel(\rho)=m\} = \mathsf{Poisson}(m;r')
\end{equation}
with $r' = r(1-\extinct{\rho})$. 
\item If $\rootdist(n)=\mathsf{NegativeBinomial}(n;\theta,q)$, then
\begin{equation}\label{eq:likelihood.finite.negbinom}
\PROB\{\snodelabel(\rho)=m\} = \mathsf{NegativeBinomial}(m;\theta,q')
\end{equation}
with $q' = \frac{q(1-\extinct{\rho})}{1-q\extinct{\rho}}$. 
\item If $\nodelabel(\rho)$ has a Bernoulli distribution, i.e., if $\rootdist(0)=1-\rootdist(1)=1-p$, then 
\begin{equation}\label{eq:likelihood.finite.Bernoulli}
\likelihoodsym = \likelihood{\rho}{0} + p\bigl(1-\extinct{\rho}\bigr) \bigl(\likelihood{\rho}{1}-\likelihood{\rho}{0}\bigr)
\end{equation}
\end{enumerate}
\end{subequations}
\end{theorem}
Consequently, the likelihood for a Poisson distribution at the root is computed as 
\begin{equation}\label{eq:likelihood.Poisson}
	\likelihoodsym=\sum_{m=0}^{M_{\rho}} \likelihood{\rho}{m} \mathsf{Poisson}\bigl(m;\Gamma(1-\extinct{\rho})\bigr),
\end{equation}
where~$\Gamma$ is the mean family size at the root.  

\section{Algorithm}
The algorithm we describe computes the likelihood 
of a phylogenetic profile for a given set of model parameters.
Algorithm \Algo{ComputeConditionals} below proceeds by  postorder (depth-first) traversals; 
the necessary variables are calculated 
from the leaves towards the root. The loop of Line~\ref{line:loop.transition} 
computes the transition probabilities $\psumei{\cdot}{\cdot}{\cdot}$, 
extinction probabilities $\extinct{\cdot}$ and survival count ranges $M_{\cdot}$. 
The loop of Line~\ref{line:loop.likelihood} carries out the computations suggested by 
Theorem~\ref{tm:likelihood.recursion}.

\begin{theorem}\label{tm:runtime}
Let~$\tree$ be a phylogeny with~$n$ nodes
where every node has at most $\numchildren^*$ children. 
Let~$h$ denote the tree height, i.e., 
	the maximum number of edges from the root to a leaf
The \Algo{ComputeConditionals} algorithm computes the 
	conditional survival likelihoods for a phylogenetic profile~$\profile$ 
	on~$\tree$
	in $O\bigl(M^2h + \numchildren^*(Mh+n)\bigr)$ time, where~$M=M_{\rho} =\sum_x \profile(x)$ is 
	the total number of homologs.
\end{theorem}

If~$\numchildren^{*}$ is constant, then the running time bound of 
Theorem~\ref{tm:runtime} is $O(M^2h+n)$. For almost all phylogenies 
in a Yule-Harding random model, $h=O(\log n)$, so the typical 
running time is~$O(M^2\log n)$. For all phylogenies, $h\le n-1$, 
which yields a~$O(M^2n)$ worst-case bound. 

\begin{center}
\footnotesize
\begin{fmpage}{0.98\textwidth}
\begin{algolist}{}
\item[] \Algo{ComputeConditionals}
\item[] \textbf{Input:} phylogenetic profile $\profile$ 
\item \textbf{for} each node $x\in\mathcal{V}(\tree)$ in 
	a postorder traversal \textbf{do}\label{line:loop.transition}
\item\mt\ Compute the sum of gene counts $M_x$ by~\eqref{eq:range}.\label{line:ranges}
\item\mt\ Compute $\extinct{x}$ using \eqref{eq:extinct}.\label{line:extinct}
\item\mt\ \textbf{if} $x$ is not the root \label{line:survive.loop}
\item\mtt\ \textbf{then} let $y$ be the parent of $x$.\label{line:survive.body}
\item\mtt\ \textbf{for} $n=0,\dotsc,M_x$ \textbf{do} 
\item\mttt\ \textbf{for} $m=0,\dotsc,M_x$ \textbf{do} 
\item\mtttt\ compute $\psumei{yx}{n}{m}$ using \eqref{eq:sumi}
	with $\ehori{x}{\cdot}$ and $\ecopies{x}{\cdot}$ 
	from Lemmas~\ref{lm:xenolog.survive} and~\ref{lm:paralog.survive}.\label{line:sumi}
\item \textbf{for} each node $x\in\mathcal{V}(\tree)$ in 
	a postorder traversal \textbf{do}\label{line:loop.likelihood}
\item\mt\ \textbf{if} $x$ is a leaf 
\item\mt\ \textbf{then} \textbf{for} all $n\setto 0,\dotsc,\profile(x)$ \textbf{do} set $\likelihood{x}{n}\setto \{n=\profile(x)\}$ 
\item\mt\ \textbf{else} 	
\item\mtt\ Let $x_1,\dotsc,x_{\numchildren}$ be the children of $x$
\item\mtt\ Initialize $M[0]\setto 0$ and $D[0]\setto 1$
\item\mtt\ \textbf{for} $i\setto 1, \dotsc, \numchildren$ \textbf{do}\label{line:loop.lineages}
\item\mttt\ set $M[i]\setto M[i-1]+M_{x_i}$ and $D[i]\setto D[i-1] \cdot \extinct{x_i}$
\item\mttt\ \textbf{for} all $t\setto 0,\dotsc, M[i-1]$ and $s\setto 0,\dotsc,M_{x_i}$ \textbf{do}
\item\mtttt\ compute $\lineagelik{i}{t}{s}$ by Eqs.~\eqref{eq:lineagelik}
\item\mttt\ \textbf{if} $i=1$ \textbf{then} \textbf{for} all $n\setto 0,\dotsc,M[i]$ \textbf{do}
	set $\sectionlik{1}{n}\setto (1-\prodextinct{1})^{-n} \lineagelik{1}{0}{n}$
\item\mttt\ \textbf{else} 
\item\mtttt\ \textbf{for} all $n\setto 0,\dotsc,M[i]$ \textbf{do} initialize $\sectionlik{i}{n}\setto 0$ 
\item\mtttt\ \textbf{for} $t\setto 0,\dotsc, M[i-1]$ and $s\setto 0,\dotsc,M_{x_i}$ \textbf{do}  
\item\mttttt\  set $\sectionlik{i}{n}\setto\sectionlik{i}{n}+\mathsf{Binomial}(s; n, \prodextinct{i-1})
		\sectionlik{i-1}{t} \lineagelik{i}{t}{s}$
\item\mtttt\ \textbf{for} all $n\setto 0,\dotsc,M[i]$ \textbf{do} $\sectionlik{i}{n}\setto \bigl(1-\prodextinct{i}\bigr)^{-n} \sectionlik{i}{n}$ 
\item\mtt\ \textbf{for} all $n\setto 0,\dotsc, M_x$ \textbf{do} set $\likelihood{x}{n} \setto \sectionlik{\numchildren}{n}$.
\end{algolist}
\end{fmpage}
\end{center}

\section{Mathematical proofs}
\begin{proof}[Proof of Lemma~\ref{lm:psum}]
Equations~\eqref{eq:sumo.0m} and~\eqref{eq:sumi.0m} are immediate 
since
\[
\psumo{0}{m} = \psumi{0}{m}=\PROB\{\xenosize=m\} = \hori{m}.
\]
By the independence of~$\parasize_i$, for all $n>0$,
\[
\psumo{n}{0} = \PROB\{\xenosize+\sum_{i=1}^n \parasize_i=0\} 
	= \psumo{n-1}{0} \PROB\{\parasize_{n}=0\} = \psumo{n-1}{0}\cdot p,
\]
as in~\eqref{eq:sumo.n0}.

Let~$\copies{n} = \mathsf{ShiftedGeometric}(n;p,q)$ be the 
common probability mass function of~$\parasize_i$. 
For~$m,n>0$, 
\begin{multline}\label{eq:sumo.sumG}
\psumo{n}{m} = \PROB\Bigl\{\xenosize+\sum_{i=1}^n \parasize_i=m\Bigr\} 
			  = \sum_{k=0}^m \PROB\{\parasize_n=k\}\cdot\PROB\Bigl\{\xenosize+\sum_{i=1}^{n-1}\parasize_i = m-k\Bigr\}\\
			  = \sum_{k=0}^m \copies{k}\cdot \psumo{n-1}{m-k}.
\end{multline}
For~$m=1$, \eqref{eq:sumo.sumG} is tantamount to~\eqref{eq:sumo.n1}, since
$\copies{0}=p$ and $\copies{1}=(1-p)(1-q)$. For~$m>1$, \eqref{eq:sumo.sumG} can be 
further rewritten using~$\copies{k}=q \copies{k-1}$ for all~$k>1$:
\begin{align*}
\psumo{n}{m} 
& = 
	\begin{aligned}[t]
		& \copies{0}\cdot \psumo{n-1}{m} + \copies{1}\cdot \psumo{n-1}{m-1}\\
		& + \sum_{k=2}^m q \copies{k-1} \cdot\psumo{n-1}{m-k}
	\end{aligned}\\
& = 	
	\begin{aligned}[t]
		& p \cdot\psumo{n-1}{m} + \copies{1}\cdot \psumo{n-1}{m-1}\\
		& + q \sum_{k=1}^{m-1} \copies{k} \cdot\psumo{n-1}{m-1-k}
	\end{aligned}\\
& = 	
	\begin{aligned}[t]
		& p \cdot\psumo{n-1}{m} + \copies{1}\cdot \psumo{n-1}{m-1}\\
		& + q \Bigl(\psumo{n}{m-1}-\copies{0}\cdot\psumo{n-1}{m-1}\Bigr), 
	\end{aligned}
\end{align*}
which leads to the recursion of~\eqref{eq:sumo.nm} since~$\copies{1}-q\copies{0} = 1-p-q$. 

Equation~\eqref{eq:sumi.nn} follows from
\[
\psumi{n}{n} = \PROB\Bigl\{\xenosize+\sum_{i=1}^n \parasize_i=n; \forall \parasize_i>0\Bigr\}
	= \PROB\{\xenosize=0\}\cdot \prod_{i=1}^n \PROB\{\parasize_i=1\} = H(0)\bigl(G(1)\bigr)^n. 
\]
For $m>n>0$, 
\begin{align*}
\psumi{n}{m} & = \PROB\Bigl\{\xenosize+\sum_{i=1}^n \parasize_i=m; \parasize_i>0\Bigr\} \\
			 & = \sum_{k=1}^{m-n+1} \PROB\{\parasize_n=k\}\cdot \psumi{n-1}{m-k}\\
			 & = \sum_{k=1}^{m-n+1} \copies{k}\cdot \psumi{n-1}{m-k}\\
			 & = \copies{1}\cdot\psumi{n-1}{m-1} + \sum_{k=2}^{m-n+1} q \copies{k-1}\cdot \psumi{n-1}{m-k}\\
			 & = \copies{1}\cdot\psumi{n-1}{m-1} + q\cdot \psumi{n-1}{m},
\end{align*}
as claimed in~\eqref{eq:sumi.nm}.
\end{proof}

Lemmas~\ref{lm:paralog.survive} and~\ref{lm:xenolog.survive} 
rely on the following general result.
\begin{lemma}\label{lm:extinctionsum.generating}
\begin{subequations}
Let $\sigma\in\mathbb{R}$ be a fixed parameter, 
and let $\{a_n\}_{n=0}^{\infty}$ and $\{b_n\}_{n=0}^{\infty}$ be two number sequences related by the formula
\begin{equation}\label{eq:extinctionsum.infinite}
b_n = \sum_{i=0}^\infty \binom{n+i}{i} \sigma^i (1-\sigma)^{n} a_{n+i}. 
\end{equation}
(We use the convention $0^0=1$ in the formula when $\sigma\in\{0,1\}$.)
Let~$A(z)=\sum_n a_n z^n$ and $B(z)=\sum_n b_n z^n$ denote the 
generating functions for the sequences. Then
\begin{equation}\label{eq:extinctionsum.generating}
B(z) = A\bigl(\sigma+(1-\sigma)z\bigr)
\end{equation}
\end{subequations}
\end{lemma}
\begin{proof}
If $\sigma=0$, then $b_n=a_n$, and, thus~\eqref{eq:extinctionsum.generating} holds.
If $\sigma=1$, then $b_0=\sum_{k=0}^\infty a_k$, and $b_n=0$ for $n>0$, 
which implies~\eqref{eq:extinctionsum.generating}. 
Otherwise,
\begin{align*}
B(z) & = \sum_{n=0}^\infty z^n \sum_{m=n}^{\infty} a_m \binom{m}{n} \sigma^{m-n} (1-\sigma)^n \\*
	& = \sum_{m=0}^\infty a_m \sum_{n=0}^m z^n \binom{m}{n} \sigma^{m-n} (1-\sigma)^n \\*
	& = \sum_{m=0}^\infty a_m \bigl(\sigma+(1-\sigma)z\bigr)^m = A\bigl(\sigma+(1-\sigma)z\bigr), 
\end{align*}
as claimed.
\end{proof}

\begin{corollary}\label{cor:extinctionsum.distribution}
Let~$\{a_n\}_{n=0}^{\infty}$ and~$\{b_n\}_{n=0}^{\infty}$ be two
probability mass functions for non-negative integer random 
variables, related as in~\eqref{eq:extinctionsum.infinite}.
\begin{enumerate}

\item If $a_n = \mathsf{ShiftedGeometric}(n;p,q)$, then 
$b_n=\mathsf{ShiftedGeometric}(n;p',q')$ with
\begin{equation}\label{eq:extinctionsum.distribution.pq}
p' = \frac{p(1-\sigma)+(1-q)\sigma}{1-q\sigma}\quad\text{and}\quad
q' = \frac{q(1-\sigma)}{1-q\sigma}.
\end{equation}

\item If $a_n=\mathsf{NegativeBinomial}(n; \theta, q)$, 
then $b_n = \mathsf{Negativebinomial}(n; \theta, q')$,
where $q'$ is defined as in~\eqref{eq:extinctionsum.distribution.pq}.

\item 
If $a_n=\mathsf{Poisson}(n;r)$, then 
$b_n = \mathsf{Poisson}(n;r')$ with 
$r'=r(1-\sigma)$.
\end{enumerate}
\end{corollary}
\begin{proof}
The corollary follows for plugging 
into Lemma~\ref{lm:extinctionsum.generating} 
the generating functions
$A(z) = \frac{p+z(1-p-q)}{1-qz}$, 
$A(z) = \bigl(\frac{1-q}{1-qz}\bigr)^{\theta}$
and
$A(z) = e^{r(z-1)}$ 
for the shifted geometric, negative binomial,
and Poisson distributions, respectively. 
\end{proof}

\begin{proof}[Proof of Lemma~\ref{lm:paralog.survive}]
Let~$\parasize$ be the random variable 
that is the size of an inparalog group at node~$y$.
In order to have~$n$ surviving inparalogs, there must be 
$n+i$ inparalogs in total, out of which~$i$ do not survive, for some~$i\ge 0$. 
Therefore,
\[
\ecopies{y}{n}
	= \sum_{i=0}^{\infty} 
 \PROB\bigl\{\parasize = n+i\bigr\} \binom{n+i}{i} \extinct{y}^i (1-\extinct{y})^n. 
\]
By Table~\ref{tbl:bd}, $\PROB\bigl\{\parasize=n\bigr\}=\mathsf{ShiftedGeometric}(n;p,q)$ 
where the distribution parameters~$p$ and~$q$ are determined by 
the parameters of the birth-and-death process on the edge leading to~$y$
($q=0$ in the degenerate case where duplication rate is~0).
The lemma thus follows from Corollary~\ref{cor:extinctionsum.distribution}
with~$\sigma = \extinct{y}$.
\end{proof}

\begin{proof}[Proof of Lemma~\ref{lm:xenolog.survive}]
Let~$\xenosize$ be the random variable that is the size of 
the xenolog group at~$y$.
In order to have~$n$ surviving xenologs, there must be 
$n+i$ xenologs in total, out of which~$i$ do not survive, for some~$i\ge 0$. 
Therefore,
\begin{equation}\label{eq:sumehori.1}
\ehori{y}{n} = \sum_{i=0}^{\infty} 
 \PROB\bigl\{\xenosize=n+i\bigr\} \binom{n+i}{i} \extinct{y}^i (1-\extinct{y})^n. 
\end{equation}
By Table~\ref{tbl:bd}
if $\drate=0$, then $\xenosize$ has a Poisson distribution with parameter~$r$;
otherwise, $\xenosize$ has a negative binomial distribution with parameters~$\theta$ and~$q$. 
In either case, the lemma follows from Corollary~\ref{cor:extinctionsum.distribution}
after setting~$\sigma=\extinct{y}$.
\end{proof}
%

\begin{proof}[Proof of Lemma~\ref{lm:extinct}]
For leaves, the statement is trivial. 
When~$x$ is not a leaf, the lemma follows 
from the fact that survivals are independent between 
disjoint subtrees.
\end{proof}

\begin{proof}[Proof of Theorem~\ref{tm:likelihood.recursion}]
The formulas are obtained by tracking survival within the lineages~$xx_i$  
among the individuals at~$x$. 
Define the indicator variables~$X_{i,j}$ for each individual 
$j=1,\dotsc,\nodelabel(x)$ and lineage $i=1,\dotsc, c$, taking the value~1
if and only if individual~$j$ has at least one surviving offspring at~$x_i$. 
In order to work with the survivals, we introduce some
auxiliary random variables that have the distributions of surviving xenologs and 
inparalogs. 
For every edge~$xx_i$, define the sequence of independent 
random variables $\bigl(\parasize_{ij}\colon j=1,2,\dotsc\bigr)$
with identical distributions~$\ecopies{x_i}{\cdot}$, and 
the random variable~$\xenosize_i$ with distribution $\ehori{x_i}{\cdot}$. 
Consequently, 
$\PROB\bigl\{X_{i,j}=0\bigr\} = \PROB\bigl\{\parasize_{ij}=0\bigr\} = \ecopies{x_i}{0}$, 
and $\Probcmd{\snodelabel(x_i)=k}{\nodelabel(x)=n} = \PROB\Bigl\{\bigl(\xenosize_i+\sum_{j=1}^{n} \parasize_{ij}\bigr)=k\Bigr\}$. 
By Lemma~\ref{lm:paralog.survive},
$\ecopies{x_i}{k} = \mathsf{ShiftedGeometric}(k;p',q')$ with some~$p'$ and~$q'$.

Define the shorthand notation 
$\profile_i=\Bigl\{\nodelabel\bigl(\tree_{x_i}\bigr)=\profile\bigl(\tree_{x_i}\bigr)\Bigr\}$
for observing the counts in the subtree rooted at~$x_i$.
Let
\[
\lineagelik{i}{t}{s} = \Probcmd{\profile_i; \sum_{j=1}^s X_{i,j}=s}{\nodelabel(x)=t+s}.
\]
In other words, if there are $t+s$ individuals at~$x$, then 
$\lineagelik{i}{t}{s}$ is the probability that~$s$ selected individuals 
survive in the lineage~$xx_i$ and the given profile is observed in~$\tree_{x_i}$.
By Lemma~\ref{lm:psum},
\begin{align*}
\lineagelik{i}{0}{s} 
	& = \Probcmd{\profile_i;
		\sum_{j=1}^s X_{i,j}=s}{\nodelabel(x)=s}\\*
	& = \sum_{m=0}^{M_{x_i}} 
		\Probcmd{\profile_i;
			\snodelabel(x_i)=m;  \sum_{j=1}^s X_{i,j}=s}{\nodelabel(x)=s}\\
	& = \sum_m \Probcmd{\profile_i}{\snodelabel(x_i)=m;\nodelabel(x)=s} 
		\cdot \Probcmd{\snodelabel(x_i)=m; \sum_{j=1}^s X_{i,j}=s}{\nodelabel(x)=s}\\
	& = \sum_m \Probcmd{\profile_i}{\snodelabel(x_i)=m}
		\cdot \PROB\biggl\{\Bigl(\xenosize_i+\sum_{j=1}^{s} \parasize_{ij}\Bigr)=m; \forall j\le s \colon \parasize_{ij}\ne 0\biggr\}\\
	& = \sum_m \likelihood{x_i}{m}\cdot \psumei{x_i}{s}{m}
\end{align*}
as shown in~\eqref{eq:llik.i0s}. 
If $t>0$, then 
\begin{align*}
\lineagelik{i}{t}{s} & = \begin{aligned}[t]
	& \Probcmd{\profile_i;
	X_{i,s+1}=0;  \sum_{j=1}^s X_{ij}=s}{\nodelabel(x)=t+s}\\*
	+ & \Probcmd{\profile_i;
	X_{i,s+1}=1; \sum_{j=1}^s X_{ij}=s}{\nodelabel(x)=t+s}
	\end{aligned}\\
	& = \begin{aligned}[t]
		& \ecopies{x_i}{0} \cdot\Probcmd{\profile_i;
	     \sum_{j=1}^s X_{ij}=s}{\nodelabel(x)=t+s-1}\\*
	    + &  \Probcmd{\profile_i;
	     \sum_{j=1}^{s+1} X_{ij}=s+1}{\nodelabel(x)=t+s}
	    \end{aligned} \\
	 & = \copies{0} \cdot \lineagelik{i}{t-1}{s} + \lineagelik{i}{t-1}{s+1},
\end{align*}
which is tantamount to~\eqref{eq:llik.its}. Equation~\eqref{eq:llik.itm} follows
from the fact that $\lineagelik{i}{t}{s}=0$ for $s>M_{x_i}$. 

For every~$i$, let~$\mathcal{Y}_i$ denote the set of individuals at~$x$ 
that survive in at least one of the lineages $x_1,\dotsc,x_i$, i.e., 
$\mathcal{Y}_i = \Bigl\{j\colon \bigl\{X_{1j}+\dotsm+X_{ij} \ne 0 \bigr\}\Bigr\}$. 
Given the exchangeability of individuals, 
$\Probcmd{\profile_1;\dotsc;\profile_i}{\mathcal{Y}_i = \mathcal{Y}}$
is the same for all sets of the same size $|\mathcal{Y}|=n$.
Let
\[
\sectionlik{i}{n} = \Probcmd{\profile_1;\dotsc;\profile_i}{|\mathcal{Y}_i| = n}.
\]
In particular, $|\mathcal{Y}_c| = \snodelabel(x)$ and, thus, $\sectionlik{c}{n} = \likelihood{x}{n}$. 

Since $\sectionlik{i}{n}$ is the same for all values of~$\nodelabel(x)\ge n$, and
$\profile_1$ is determined solely by survival in lineage~$xx_1$, 
\begin{align*}
\lineagelik{1}{0}{n} 
	& = \Probcmd{\profile_1;
		\sum_{j=1}^n X_{i,j}=n}{\nodelabel(x)=n}\\*
	& = \Probcmd{\sum_{j=1}^n X_{i,j}=n}{\nodelabel(x)=n} 
		\cdot \Probcmd{\profile_1}{\sum_{j=1}^n X_{i,j}=n; \nodelabel(x)=n}\\
	& = (1-\prodextinct{1})^n \Probcmd{\profile_1}{|\mathcal{Y}_1|=n} = (1-\prodextinct{1})^n \sectionlik{1}{n},
\end{align*}
implying~\eqref{eq:recursion.section.1}.

By a similar reasoning for~$i>1$, 
\begin{multline}\label{eq:rspi}
\Probcmd{\profile_1;\dotsc;\profile_i; |\mathcal{Y}_i|=n}{\nodelabel(x)=n}\\
	\begin{aligned}[t] 
		& = \Probcmd{|\mathcal{Y}_i|=n}{\nodelabel(x)=n} \cdot \Probcmd{\profile_1;\dotsc;\profile_i}{|\mathcal{Y}_i|=n;\nodelabel(x)=n}\\*
		& = (1-\prodextinct{i})^n \sectionlik{i}{n}
	\end{aligned}
\end{multline}
Let~$\mathcal{X}_i$ denote the set of individuals that survive in the lineage~$xx_i$: 
$\mathcal{X}_i = \bigl\{j\colon X_{i,j}\ne 0\bigr\}$. 
We rewrite the left-hand side of~\eqref{eq:rspi} by conditioning 
on the set of individuals that survive in lineage~$xx_i$ but not in 
the lineages~$xx_1$,\ldots,$xx_{i-1}$.
\begin{multline*}
\Probcmd{\profile_1;\dotsc;\profile_i; |\mathcal{Y}_i|=n}{\nodelabel(x)=n}\\
	\begin{aligned}[t]
	 & = 
		\sum_{\mathcal{S}\in 2^{[n]}} \begin{aligned}[t]
			& \Probcmd{\profile_i; \mathcal{X}_i\setminus\mathcal{Y}_{i-1}=\mathcal{S}}{\nodelabel(x)=n} \\
			&\times\Probcmd{\profile_1;\dotsc;\profile_{i-1}; \mathcal{Y}_{i-1}=[n]\setminus\mathcal{S}}{\nodelabel(x)=n}
		\end{aligned}\\
	& = \sum_{s+t=n} \lineagelik{i}{t}{s} \sectionlik{i-1}{t} \Probcmd{|\mathcal{Y}_{i-1}|=t}{\nodelabel(x)=n}\\
	& = \sum_{s+t=n} \lineagelik{i}{t}{s} \sectionlik{i-1}{t} \binom{t+s}{s} (\prodextinct{i-1})^s(1-\prodextinct{i-1})^t. 
	\end{aligned}
\end{multline*}
Combining this latter equality with~\eqref{eq:rspi} leads to~\eqref{eq:recursion.section.i}.
\end{proof}

\begin{proof}[Proof of Theorem~\ref{tm:likelihood.finite}]
By~\eqref{eq:snodelabel.calc}, Lemma~\ref{lm:extinctionsum.generating}
with~$\sigma=\extinct{x}$
shows the relationship between the generating functions for the distributions
of~$\nodelabel(x)$ and~$\snodelabel(x)$. The theorem considers 
the case when~$x$ is the root and Corollary~\ref{cor:extinctionsum.distribution}
applies to~$\rootdist(n)=\PROB\{\nodelabel(x)=n\}$.
\end{proof}

\begin{proof}[Proof of Theorem~\ref{tm:runtime}]
By Theorem~\ref{tm:likelihood.recursion}, the algorithm correctly computes the conditional survival likelihoods. 
Let~$\tree$ be the phylogeny with root~$\rho$ and~$n$ nodes. 
In order to prove the running time result, consider first the loop of Line~\ref{line:loop.transition}. 
Lines~\ref{line:ranges}--\ref{line:survive.body} take~$O(1)$ time for each~$x\in\mathcal{V}(\tree)$. 
Line~\ref{line:sumi} is executed~$(M_x+1)^2$ times for every non-root~$x$. 
If~$x$ is an inner node with children~$x_1,x_2,\cdots,x_{\numchildren}$, 
then $M_x=\sum_{j=1}^{\numchildren} M_{x_j}$. Consequently, 
\begin{equation}\label{eq:range.recursion.bound}
\sum_{j=1}^{\numchildren} (M_{x_j}+1)^2 \le (M_x+1)^2 + (\numchildren-1).
\end{equation}
Now, consider the {\em tree levels} $\mathcal{V}_0,\mathcal{V}_h$, 
where~$\mathcal{V}_0=\{\rho\}$, and for all~$i=1,\dotsc,h$,
$\mathcal{V}_i$ consists of all the children of nodes in~$\mathcal{V}_{i-1}$. 
In other words, $\mathcal{V}_i$ is the set of nodes that are reached through~$i$ edges 
from the root. By~\eqref{eq:range.recursion.bound},
\[
\sum_{y \in \mathcal{V}_i} (M_y+1)^2 
	\le \bigl|\mathcal{V}_i\bigr|-\bigl|\mathcal{V}_{i-1}\bigr|+\sum_{x\in\mathcal{V}_{i-1}} (M_x+1)^2. 
\]
for all~$i>0$. Therefore,
\begin{equation}\label{eq:range.levels.bound}
\sum_{y \in \mathcal{V}_i} (M_y+1)^2 \le \Bigl(\bigl|\mathcal{V}_i\bigr|-1\Bigr) + (M_{\rho}+1)^2
\end{equation}
So,
\begin{align*}
\sum_{x\in \mathcal{V}(\tree)\setminus\{\rho\}} (M_x+1)^2
	& = \sum_{i=1}^h \sum_{x\in\mathcal{V}_i} (M_x+1)^2 \\
	& \le n-1-h + h (M_{\rho}+1)^2\\
	& = O\bigl(n+hM^2\bigr).
\end{align*}
Therefore, executing Line~\ref{line:sumi} through all iterations takes~$O\bigl(M^2h+n)$ time. 
In order to bound the loop's running time in Line~\ref{line:loop.likelihood}, 
consider the $\lineagelik{i}{t}{s}$ and $\sectionlik{i}{n}$ values
that are needed for a given node~$x$ with children $x_1,\dotsc,x_{\numchildren}$. 
By~\eqref{eq:llik.i0s}, computing all~$\lineagelik{i}{0}{s}$ values takes 
$O((M_{x_i}+1)^2)$ time. Every $\lineagelik{i}{t}{s}$ with $t>0$ and 
$\sectionlik{i}{n}$ is calculated in~$O(1)$ time. Using the bound~$M[i]\le M_x$, 
iteration~$i$ of the loop in Line~\ref{line:loop.lineages} takes $O\bigl((M_x+1)(M_{x_i}+1)\bigr)$ 
time. By summing for~$i=1,\dotsc,\numchildren$, we get that for node~$x$ with~$\numchildren_x$
children, the loop of Line~\ref{line:loop.likelihood} takes $O\bigl((M_x+1)(M_x+\numchildren_x)\bigr)$
time (since $\sum_i (M_{x_i}+1)=M_x+\numchildren$). 
Now, $(M_x+1)(M_x+\numchildren_x) = (M_x+1)(M_x+1+\numchildren-1)$, and 
\begin{align*}
\sum_{x\in\mathcal{V}(\tree)} (M_x+1)(\numchildren_x-1) 
	& = \sum_{i=0}^h \sum_{x\in\mathcal{V}_i} (M_x+1)(\numchildren_x-1) \\
	& \le \bigl(\max_x \numchildren_x-1\bigr) \sum_{i=0}^h \sum_{x\in\mathcal{V}_i} (M_x+1)\\*
	& \le \bigl(\numchildren^{*}-1\bigr) (M_{\rho} (h+1)+n).
\end{align*}
By our previous discussions, 
\begin{align*}
\sum_{x\in\mathcal{V}(\tree)} (M_x+1)(M_x+\numchildren_x) 
	& \le n-1-h + h (M_{\rho}+1)^2 + (\numchildren^*-1) (M_{\rho} (h+1)+n)\\
	& = O\bigl(M^2h + \numchildren^*(Mh+n)\bigr).
\end{align*}
So, the loop of Line~\ref{line:loop.likelihood} takes 
$O(M^2h + Mh\numchildren^*)$ time, which leads to the Theorem's claim
when combined with the bound on the 
loop's running time in Line~\ref{line:loop.transition}. 
\end{proof}

\section{Likelihood correction for absent profiles} 
Suppose that profiles are restricted to the 
condition that $\bigl\{\profile(x)>0\bigr\}$ must hold for at least one terminal node~$x$.
The corresponding likelihoods 
\[
\likelihoodsym_1 = \Probcmd{\forall x\in\leaves{\tree}\colon \nodelabel(x)=\profile(x)}{\text{$\nodelabel(x)>0$ for at least one leaf}} 
\]
are obtained from the full likelihood by 
employing a correction that involves the probability of the condition~\citep{Felsenstein.restml}. 
Namely,
\[
\likelihoodsym_1 = \frac{\likelihoodsym}{\PROB\Bigl\{\text{$\nodelabel(x)>0$ for at least one leaf}\Bigr\}}
	= \frac{\likelihoodsym}{1-\PROB\Bigl\{\text{$\nodelabel(x)=0$ for all leaves $x$}\Bigr\}}. 
\]
The probability that 
$\nodelabel(x)=0$ at all the leaves is the 
likelihood of the all-0 profile $\profile_0 = (0,\dotsc,0)$. By Theorem~\ref{tm:likelihood.recursion}, 
$\likelihood{\rho}{0}=\prod_{xy\in\mathcal{E}(\tree)} \ehori{y}{0}$ for the profile~$\profile_0$. 
Combined with~\eqref{eq:likelihood.Poisson}, 
we have the correction formula
$\likelihoodsym_1 = \frac{\likelihoodsym}{1-p_0}$
with 
\begin{equation}\label{eq:p0}
p_0=\Biggl(\,\prod_{xy\in\mathcal{E}(\tree)} \ehori{y}{0}\Biggr)\cdot
	\exp\Bigl(\Gamma(1-\extinct{\rho})\Bigr),
\end{equation}
for $\rootdist(n) = \mathsf{Poisson}(n;\Gamma)$.

\section{Inferring family sizes at ancestors and counting lineage-specific events} 
Given a profile~$\profile$, the posterior probabilities for 
gene family size at node~$x$
are computed by 
using the conditional survival likelihoods~$\likelihood{x}{n}$ 
and likelihoods of some relevant profiles on truncated phylogenies. 
In order to compute the gene content at node~$x$, for example,
consider the profile~$\profile_{x:m}$ for all~$m$ that applies to a 
phylogeny obtained by pruning the edges below~$x$, that is, 
$\profile_{x:m}(y)=\profile(y)$ for $y\not\in\tree_x$ 
and $\profile_{x:m}(x)=m$. Let~$\likelihoodsym_{x:m}$ denote the likelihood of~$\profile_{x:m}$
on the pruned tree. 
Then 
\[
\Probcmd{\nodelabel(x)=m}{\nodelabel\bigl(\leaves{\tree}\bigr)=\profile}
	= \frac{\likelihoodsym_{x:m}  \sum_{n=0}^m \binom{m}{n} (\extinct{x})^{m-n} (1-\extinct{x})^{n} \likelihood{x}{n}}
	{\likelihoodsym}
\]
gives the posterior probability that the family had~$m$ homologs at node~$x$. 
The number of families present at node~$x$, denoted by~$N_x$,  
is inferred as a posterior mean value by summing posterior probabilities:
\begin{multline}\label{eq:Nx}
N_x = \sum_{i=1}^n \Probcmd{\nodelabel(x)>0}{\nodelabel\bigl(\leaves{\tree}\bigr)=\profile_i}
	\\+ \frac{n\cdot p_0}{1-p_0} \Probcmd{\nodelabel(x)>0}{\nodelabel\bigl(\leaves{\tree}\bigr)=\profile_0},
\end{multline}
where $\profile_i\colon i=1,\dotsc,n$ are the profiles in the data set and $p_0$ is the likelihood 
of the all-0 profile~$\profile_0$ from~\eqref{eq:p0}.  Notice that the formula includes 
the absent all-0 profiles; there are $\frac{n p_0}{1-p_0}$ such profiles by expectation.

For each edge~$xy$, the posterior probabilities for {\em gain},
{\em loss}, {\em expansion} and {\em contraction} are:
\begin{align*}
\PROB\bigl\{\textrm{gain}(xy)\bigr\} & = \PROB\bigl\{\nodelabel(x)=0,\nodelabel(y)>0\bigr\} 
	 = \PROB\bigl\{\nodelabel(x)=0\bigr\}-\PROB\bigl\{\nodelabel(x)=0,\nodelabel(y)=0\bigr\}\\
\PROB\bigl\{\textrm{loss}(xy)\bigr\} & = \PROB\bigl\{\nodelabel(x)>0,\nodelabel(y)=0\bigr\} 
	 = \PROB\bigl\{\nodelabel(y)=0\bigr\}-\PROB\bigl\{\nodelabel(x)=0,\nodelabel(y)=0\bigr\}\\
\PROB\bigl\{\textrm{expansion}(xy)\bigr\} & = \PROB\bigl\{\nodelabel(x)=1,\nodelabel(y)>1\bigr\} 
	= \PROB\bigl\{\nodelabel(x)=1\bigr\}\\ &-\PROB\bigl\{\nodelabel(x)=1,\nodelabel(y)=0\bigr\}-\PROB\bigl\{\nodelabel(x)=1,\nodelabel(y)=1\bigr\}\\
\PROB\bigl\{\textrm{contraction}(xy)\bigr\} & = \PROB\bigl\{\nodelabel(x)>1,\nodelabel(y)=1\bigr\} 
	= \PROB\bigl\{\nodelabel(y)=1\bigr\}\\&-\PROB\bigl\{\nodelabel(x)=0,\nodelabel(y)=1\bigr\}-\PROB\{\nodelabel(x)=1,\nodelabel(y)=1\bigr\},
\end{align*}
where all probabilities are conditioned on the observation of the
phylogenetic profile~$\bigl\{\nodelabel\bigl(\leaves{\tree}\bigr)=\profile\bigr\}$. 
Expected numbers for gains, losses, expansions and contractions on each edge~$xy$ are computed 
by formulas analogous to~\eqref{eq:Nx}. 

\begin{figure}
\centerline{\includegraphics[width=\textwidth]{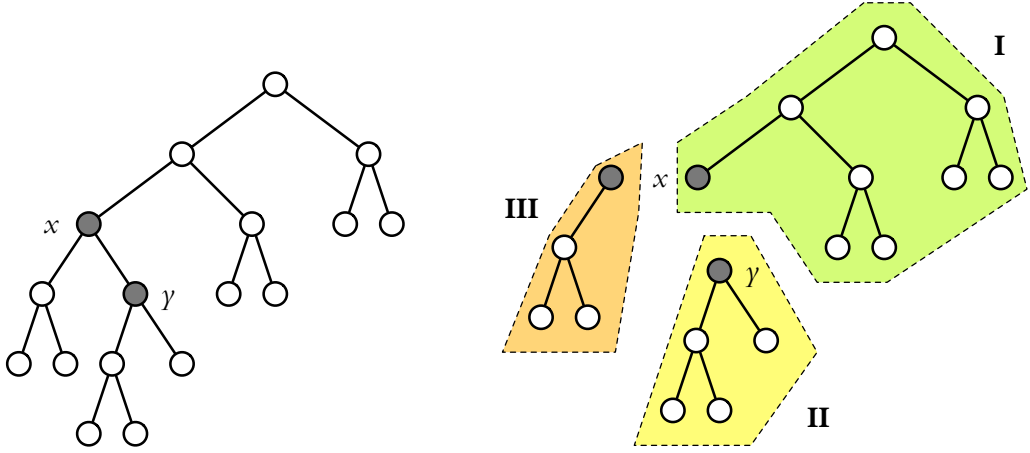}}
\caption[Tree decomposition for computing posterior probabilities]{Decomposition of the phylogeny in order to compute posterior probability for 
lineage-specific events. Equation~\eqref{eq:posterior.edges} shows the factors corresponding to the three parts.}\label{fig:posterior.decomposition}
\end{figure}

Posterior probabilities of the general form 
$\Probcmd{\nodelabel(x)=n,\nodelabel(y)=m}{\nodelabel\bigl(\leaves{\tree}\bigr)=\profile}$,
characterizing lineage-specific family size changes on edge~$xy$,
can also be computed by using survival likelihoods on truncated phylogenies.
In particular, we decompose the events as
\begin{multline}\label{eq:posterior.edges}
\PROB\Bigl\{\nodelabel(x)=n,\nodelabel(y)=m, \nodelabel\bigl(\leaves{\tree}\bigr)=\profile\Bigr\} = \textbf{I} \times \textbf{II} \times \textbf{III}\\
	= \begin{aligned}[t]
		& \PROB\Bigl\{\nodelabel(x)=n, \nodelabel\bigl(\leaves{\tree}\setminus\leaves{\tree_x}\bigr)=\profile\bigl(\leaves{\tree}\setminus\leaves{\tree_x}\bigr)\Bigr\}\\
		\times & \Probcmd{\nodelabel(y)=m, \nodelabel\bigl(\leaves{\tree_y}\bigr)=\profile\bigl(\leaves{\tree_y}\bigr)}{\nodelabel(x)=n}\\
		\times & \Probcmd{\nodelabel\bigl(\leaves{\tree_x}\setminus\leaves{\tree_y}\bigr)=\profile\bigl(\leaves{\tree_x}\setminus\leaves{\tree_y}\bigr)}{\nodelabel(x)=n},
		\end{aligned}
\end{multline}
where the second factor can be written as 
\begin{multline*}
\Probcmd{\nodelabel(y)=m, \nodelabel\bigl(\leaves{\tree_y}\bigr)=\profile\bigl(\leaves{\tree_y}\bigr)}{\nodelabel(x)=n}\\
	= \Probcmd{\nodelabel(y)=m}{\nodelabel(x)=n} \sum_{k=0}^m \binom{m}{k} (\extinct{y})^{m-k} (1-\extinct{y})^{k} 
		\likelihood{y}{k}.
\end{multline*}
Figure~\ref{fig:posterior.decomposition} illustrates the decomposition of the phylogeny into three 
parts, corresponding to the three factors in~\eqref{eq:posterior.edges}. 

\end{document}